\documentclass[final,twocolumn]{elsarticle}

\usepackage{lineno,hyperref}

\modulolinenumbers[5]

\journal{Future Generation Computer Systems}

\begin{document}

\begin{frontmatter}

\title{A Science Gateway for Exploring the X-Ray\\ Transient and Variable Sky Using EGI Federated Cloud}

\author[cnr]{Daniele D'Agostino\corref{mycorrespondingauthor}}
\author[cnr]{Luca Roverelli}
\author[cnr]{Gabriele Zereik}
\address[cnr]{CNR-Istituto di Matematica Applicata e Tecnologie Informatiche ``Enrico Magenes'', \\via Dei Marini 6, 16149 Genova, Italy }
\cortext[mycorrespondingauthor]{Corresponding author}

\author[egi]{Giuseppe La Rocca}
\address[egi]{EGI Foundation, Science Park 140 1098 XG Amsterdam The Netherlands}

\author[inaf,infn]{Andrea De Luca}
\author[inaf]{Ruben Salvaterra}
\author[inaf]{Andrea Belfiore}
\address[inaf]{INAF-Istituto di Astrofisica Spaziale e Fisica Cosmica Milano, \\via E. Bassini 15, 20133 Milano, Italy}

\author[iuss]{Gianni Lisini}
\author[iuss,inaf]{Giovanni Novara}
\author[iuss,inaf,infn]{Andrea Tiengo}
\address[iuss]{Scuola Universitaria Superiore IUSS Pavia, piazza della Vittoria 15, 27100 Pavia, Italy}

\address[infn]{Istituto Nazionale di Fisica Nucleare, Sezione di Pavia, via A. Bassi 6, 27100 Pavia, Italy}

\begin{abstract}
Modern soft X-ray observatories can yield unique insights into time domain astrophysics, and a huge amount of information is stored - and largely unexploited - in data archives.
Like a treasure-hunt, the EXTraS project harvested the hitherto unexplored temporal domain information buried in the serendipitous data collected by the European Photon Imaging Camera instrument onboard the ESA
XMM-Newton, in 16 years of observations. All results have been released to the scientific community, together with new software analysis tools.  
This paper presents the architecture of the EXTraS science gateway, that has the goal to provide the software to the scientific community through a Web based portal using the EGI Federated Cloud infrastructure. 
The main focus is on the light software architecture of the portal and on the technological insights for an effective use of the EGI ecosystem.

\end{abstract}



\end{frontmatter}


\section{Introduction}
Almost all astrophysical objects, from stars in the surroundings of the solar system, to supermassive black holes in the nuclei of very distant galaxies, display a distinctive variability as a function of time, their flux and spectral shape changing on a range of time scales. This is especially true in the high-energy range of the electromagnetic spectrum. The X-ray and gamma-ray sky is extremely dynamic and new classes of objects, some of them completely unexpected, have been discovered in the last decades thanks to their peculiar variability. Most of the variable phenomena have been discovered with large field-of-view instruments operating at hard X-ray/gamma-ray energies, which, constantly observing large portions of the sky, can also detect relatively rare events.

At soft X-rays (~0.1-10 keV), wide field instruments are much less sensitive than narrow field telescopes with focusing optics. In particular, the European Photon Imaging Camera (EPIC, \cite{strueder,turner}) instrument onboard the European Space Agency mission XMM-Newton is the most powerful tool to study the variability of faint X-ray sources, thanks to its unprecedented combination of large effective area, good angular, spectral and temporal resolution, and pretty large field of view. 


Seventeen years after its launch, EPIC is still fully operational and its immensely rich archive of data, the XMM-Newton Science Archive (XSA), keeps growing.  Large efforts are ongoing, to explore the serendipitous content in XMM data.
Indeed, the catalog of serendipitous sources extracted from EPIC observations, dubbed 3XMM, is the largest and most sensitive compilation of X-ray sources ever produced, listing more than 500,000 detections over about 800 square degrees of the sky \cite{rosen}. Further 20,000 sources have been identified in the so-called XMM Slew Survey (XSS) \cite{read}, using data collected while the telescope is moving from one target to the next one. These data present a shallower sensitivity, but they cover more than 70\% of the sky. Time-domain information on such a large sample of sources remains, however, largely unexplored 

The EU-funded FP7 Exploring the X-ray Transient and variable Sky\footnote{http://www.extras-fp7.eu}  (EXTraS) project  \cite{extras1} investigated and extracted the serendipitous content of the EPIC database in the time domain. In particular EXTraS extended 3XMM by designing and implementing four main lines of analysis:
i) a systematic study of aperiodic, short-term variability of 3XMM sources on all possible time scales (from the duration of an observation to the instrument time resolution, typically 73 ms and 2.6 sec.); ii) a systematic search for short, weak transient sources that are above the detection threshold just for a small interval of time; iii) a systematic study of long-term variability (i.e. variability between different observations), thanks to the large number of overlapping observations performed in 16 years; iv) a systematic search for periodicities.

As the most sensitive search for variability ever performed, EXTraS raises new questions in high-energy astrophysics \cite{science} and may serve as a pathfinder for future missions. Therefore EXTraS results, together with new software tools related to the four lines of analysis, has been released to the whole community.
The software is of particular importance for enhancing the potential of discovery of the XMM-Newton mission \cite{mus}, especially because it is still ongoing and fully operational and therefore it collects  new data each day.

In this paper we present the architecture of the science gateway developed in the project, named EXTraS portal\footnote{http://portal.extras-fp7.eu}, that presently allows the analysis for transient X-ray sources. 
The portal is the result of the joint effort of the two communities of the project, astrophysics and ICT, and the main contribution is represented by the description of the microservice-based approach we are following, that is based on a previous experience in deploying a science gateway for the Hydrometeorological scientific community, named DRIHM portal \cite{drihm1, drihm2}. 

The first release of the EXTraS portal has been presented in \cite{iwsg}, while in this paper we describe an improved version, based on the re-design of many components, that relies on the EGI Federated Cloud \cite{fedcloud} for executing analysis tasks.

The paper is organized as follows. Section \ref{sec2} discusses related works, Section \ref{sec3} presents the EXTraS Portal architecture while \ref{sec4} presents the EGI computational infrastructure. Section \ref{sec5} describes the service for the analysis of transient X-ray sources, while Section \ref{sec6} concludes.

\section{Related Works}
\label{sec2}

The common strategies for providing software tools to the scientific community are the following. The first, basic solution is to make available an installer or an archive containing all the files required to compile and run the analysis tool. 
This approach has been adopted for some important tools of the Astrophysics community as the Science Analysis System\footnote{http://www.cosmos.esa.int/web/xmm-newton/sas-download} (SAS), a collection of scripts and libraries specifically designed for  the XMM-Newton observations.
A second solution is to provide the software by exporting the corresponding workflows, that can be thus executed using a Workflow Management System. Also this solution has been adopted by the Astrophysics community \cite{AT, AT2}. The third solution is to provide a virtual machine with all the software installed on it.  This solution is probably the most effective for non-expert astronomers, who wants to run a few experiments, and for dissemination purposes, for example for educational programs or citizen scientists. It is worth noting that the SAS are available also as a Linux virtual machine. 
The fourth solution is to provide the software as a set of services through a Web portal designed following the science gateway paradigm \cite{sg2}. 

We adopted this strategy because science gateways are gaining an increasing interest in many communities \cite{sg}, as the astrophysics one \cite{visivo, becciani2, astro3}.
Science gateways can be defined as a set of software, data collections, instrumentations and computational capabilities that are integrated via a Web portal (or a desktop application) in a user friendly and effective environment supporting the scientific research and education activities of a specific community. 
In general, each community presents different requirements,  due to the software and/or data it shares and the goals it aims to achieve, but normally there is the need to setup services for user management, the deployment of user interfaces (UI) for experiment definition and configuration, with workflow engine and job submission services for orchestrating the experiments execution.

Some of these items can be addressed by general-purpose, ready-to-use solutions (e.g Grid certificates, workflow management systems), some others instead rely on ad-hoc solutions (i.e. the UI) that can be developed using general-purpose technologies. In general, science gateway toolkits as gUSE \cite{guse},  Apache Airavata \cite{airavata}, Agave \cite{agave} and HUBzero \cite{hubzero} are powerful solutions that allow non-ICT users to quickly deploy Web portals by providing a set of enabling technologies, front-end and back-end gateway services.  For example gUSE allows submitting jobs through the portlet technology on almost all the distributed computing infrastructures in Europe and US. Airavata can be used as a middleware layer and it provides APIs to submit and monitor jobs from gateways front-end written in several languages (e.g. Java, PHP, Python and C++). Agave focuses on providing a ``Science-as-a-Service'' solution for hybrid cloud computing, i.e. an enhancement of the ``Platform-as-a-Service'' paradigm with computational, data, and collaborative services as well. The defining characteristics of HUBzero are the delivery of visual simulation tools, that look like simple Java applets embedded within the browser, and the strong focus on the collaborative aspects for  research and educational activities.

Besides these toolkits, another interesting solution relies on Jupiter\cite{jp} and JupiterHub\cite{jph}. Jupiter is an interactive computing environment based on the concept of notebook, i.e. documents including text, plots, equations, videos but also live code and interactive widgets. In particular the output generated by running code is embedded in the notebook, which makes it a complete and self-contained record of a computation. While a Jupiter notebook runs on the workstation of a user, Jupyterhub represents a multi-user centralized environment with many software tools pre-installed, allowing scientists to access and share notebooks with a minimal effort.

However, none of these toolkits is able to provide at a time an effective solution for a) the fast development and update of fully-featured user interface also from non-IT experts; b) the exploitation of heterogeneous distributed computing infrastructures (DCI) ; c) the result sharing. Many toolkits in fact provide a native support for c) except gUSE, which however is the best solution for b) because it supports most of the DCI in Europe and in the US. As regards a), this aspect is of particular importance in designing and maintaining science gateways characterized by a continuous improvement of the software used by the community, as in our case.

Furthermore, most of these toolkits are based on time-proved technologies. These reasons, together the experience gained in refactoring  the DRIHM portal \cite{dago}, led us to investigate an alternative solution based on the PortalTS Web Portal.

\section{The EXTraS Portal Architecture}
\label{sec3}

\begin{figure*}[!hbt]
\centering
\includegraphics[width=\textwidth]{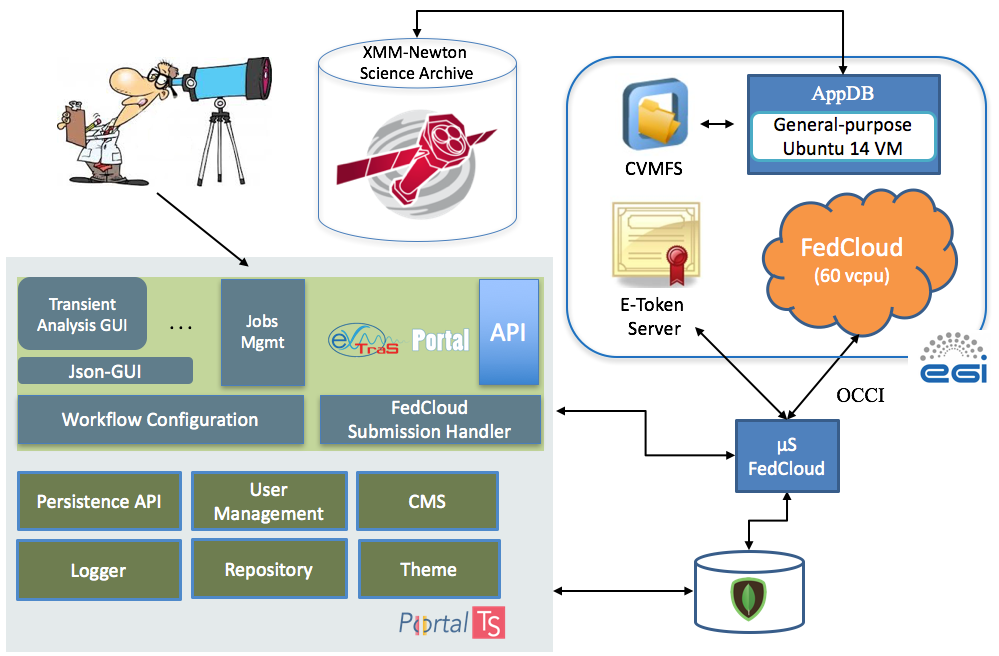}
\caption{The architecture of the EXTraS science gateway}
\label{fig_arch}
\end{figure*}


The EXTraS portal has been  developed as a set of modules of PortalTS, a modern Web Portal under development as an independent project at CNR-IMATI for the refactoring of the DRIHM portal. 
The module is a Single Page Web Application leveraging on other ready-to-use components and APIs exposed by PortalTS to enable users management, security and to persist users data.
In order to simplify the access to the FedCloud infrastructure, we also developed another component, that exposes a simple API to perform the submission and the monitoring of jobs in the Cloud. This component has been developed as a stand-alone Web Service, with its own database. For this reason, it represents a microservice.

The architecture of the EXTraS Portal and its main components are shown in Figure \ref{fig_arch}.

\subsection*{PortalTS}
PortalTS\footnote{https://portalts.it/} is an original Web Portal developed in Typescript using the NodeJS and Express frameworks. It is composed by reusable modules and implements standard features available for a Web site (e.g. user management and registration) along with other features (such as a simple API for data persistence) that enables fast development of custom modules.

A module is a component that implements and exposes a feature, but can also use features exposed by other modules. It is a very general component representing, for example, a set of web pages, a web service, a web app (aka Single Page Application), a set of static files like css files, images, or something different. According to the NodeJS philosophy, each module should be as simple as possible and implement a single functionality. 

PortalTS loads modules on the bootstrap phase, using a configuration file to ensure modules loading order.
Since PortalTS and any module are developed in Typescript, module error loading caused by typos and other typical Javascript errors (e.g. undefined functions or undefined function arguments) is dramatically reduced, resulting in an improved software reliability and stability.

The main,  ready-to-use modules PortalTS includes are the following:

The Database module defines and manages the connection with the MongoDB database. Despite its simplicity, this module is fundamental and it is used by higher-level modules for the communication with the database. 
MongoDB has been chosen because it is extremely well integrated with NodeJS. Moreover, there are some high level libraries, e.g. mongoose\footnote{http://mongoosejs.com/}, that are stable and maintained, making them usable in a production environment.

The User Management module defines an API for a complete authentication system, including user registration, login, and administration pages. It implements also role and group concepts, at the basis of the authorization mechanism for pages, modules and other entities.

The Persistence API module defines the interface to store, retrieve and manage heterogeneous data on the MongoDB database. It exposes both a RESTful and an internal API, that can be directly used  by other modules, as the CMS described below. The RESTful API is very important since it allows to store data directly from a web app, that can be built upon this modules. One example is exactly represented by the EXTraS-specific modules described below. The Persistence API defines entities and collections. A collection is a set of entities, while an entity represents possibly heterogeneous data stored with some additional metadata, like creation, update time, the owner and authorized users. An entity can belong to only one collection.

The Persistence API relies on the User Management Module to ensure security and user authentication on the data. By default, an entity is only accessible by the owner, but its read and write access policy can be changed, using a group-based policy. The Persistence API Module is fundamental since it allows to store and retrieve persistence data without any effort, enabling a ready-to-use persistence layer. Moreover, this layer is integrated with an AngularJS library that implements all the methods necessary to give a quicker and simpler access to the persistent data.

The Content Management System (CMS) module defines some web pages for user login and registration, and it allows the creation of user-defined web pages and menus. Each web page or menu element can be publicly available or accessible by a particular group, since the CMS module uses the Persistence API module. Images and files can be managed using the Repository module, and then exploited by the web pages.There are also some further basic modules, like the Theme module, that defines the web pages header and the footer to define a standard look and feel of a portal instance, and the Logging module for storing the requests received by all the modules, together with possible errors and exceptions.

\subsection*{The EXTraS Portal Modules}
\begin{figure*}[!hbt]
\centering
\includegraphics[width=\textwidth]{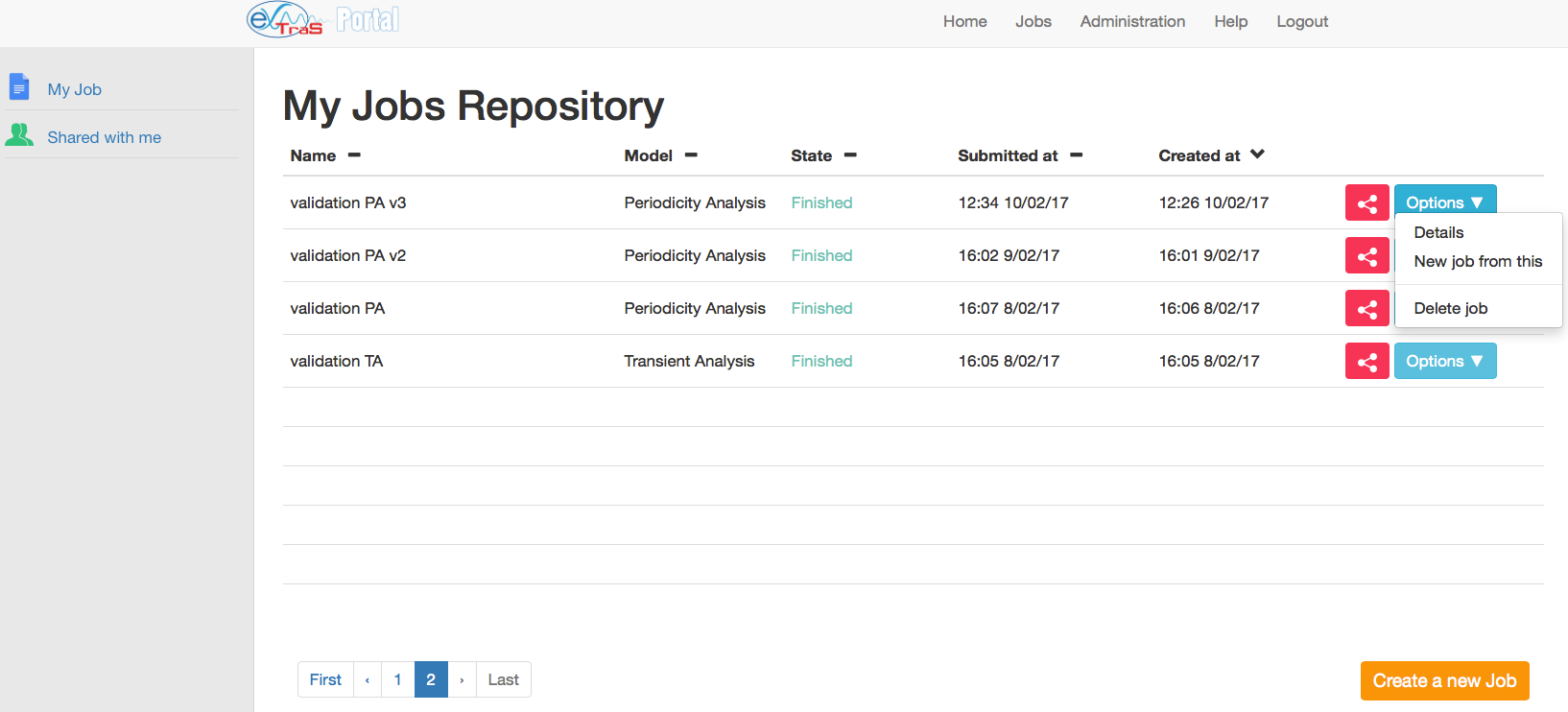}
\caption{The Jobs Management module interface.}
\label{fig_jm}
\end{figure*}

The Jobs Management module represents the Home page of the EXTraS portal, and is shown in Figure \ref{fig_jm}. It provides users with the possibility to create, submit and manage the different analysis experiments based on the software developed within the EXTraS project. In particular it presents all the submitted or configured analyses, providing the possibility to create a new analysis starting from an existing configuration or share results with other users.

This module is based on AngularJS and it is a complete web app, without any server side code. It uses the Persistence API to store and retrieve experiments data, and it activates the other portal modules corresponding to the different operations available.
With respect to the previous version of the portal in fact, the Jobs Management module has been redesigned to cooperate with other two modules, the Workflow Configuration module and the FedCloud Submission Handler module.

\begin{figure*}[!hbt]
\centering
\includegraphics[width=\textwidth]{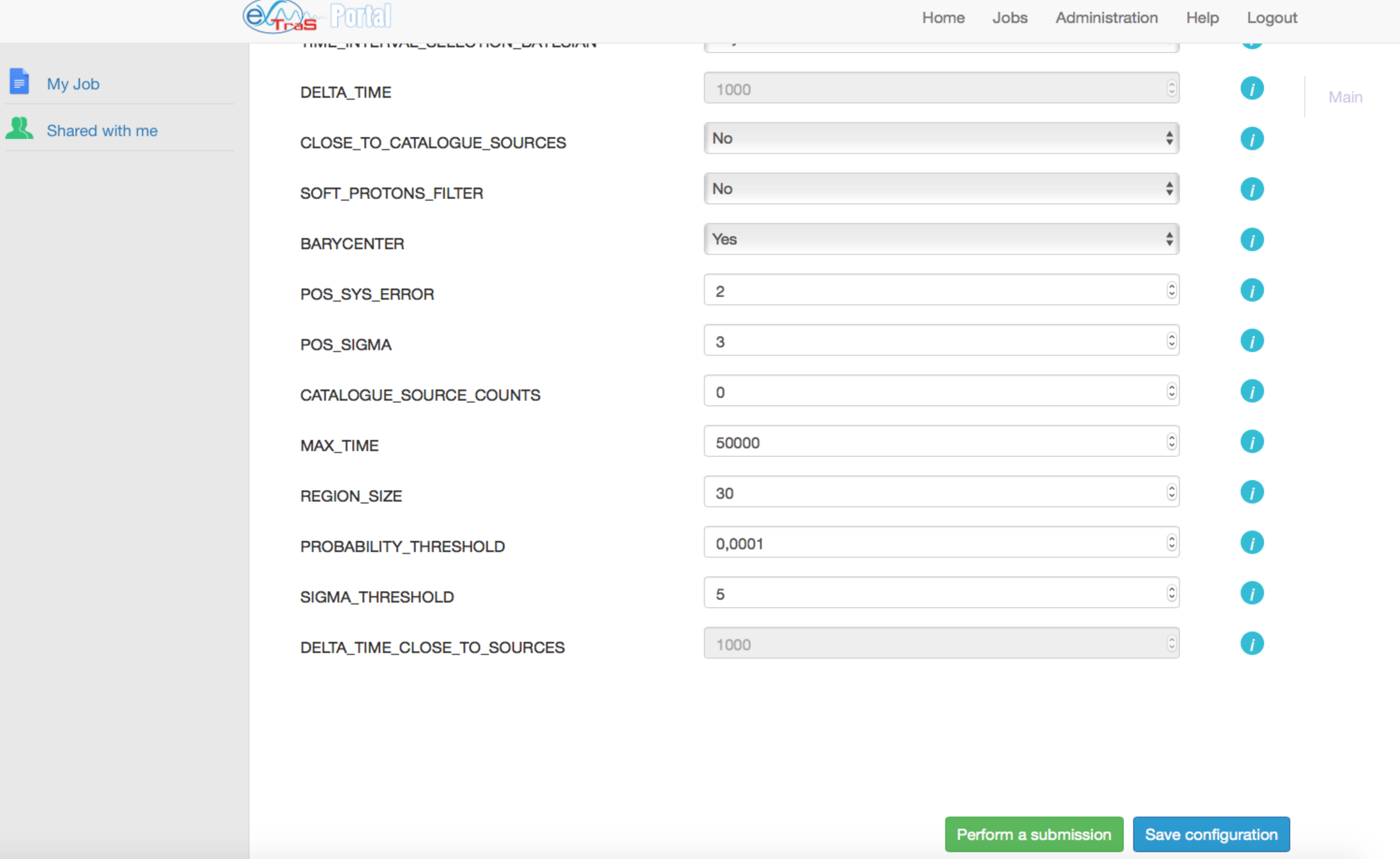}
\caption{The Transient Analysis UI shown by the Workflow configuration module.}
\label{fig_wc}
\end{figure*}
\begin{figure*}[!hbt]
\centering
\includegraphics[width=\textwidth]{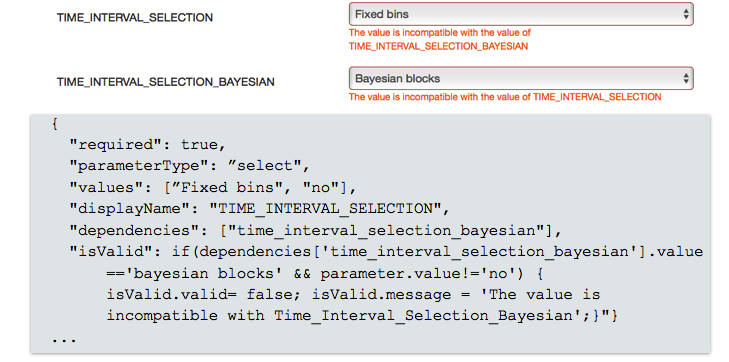}
\caption{An example of parameter and consistency check definition with Json-GUI.}
\label{fig_json}
\end{figure*}

The Workflow Configuration module is responsible to interact with the user for the creation and configuration of experiments based on the EXTraS software. In the portal every analysis corresponds to a single application, therefore there is no need to explicitly create and manage workflows. For this reason a user interacts directly with an analysis-specific UI, as shown in Figure \ref{fig_wc} for the Transient Analysis. Its main aim is to collect the parameters value and to create a namelist, that will be provided to the FedCloud Submission Handler for the actual execution of the job. A key aspect is represented by the fact that all the UI are defined as JSON files (a small example corresponding to the Transient Analysis UI is shown in Figure \ref{fig_json}) stored in the portal repository and accessed via the Persistence API. The UI corresponding to the requested analysis is therefore dynamically created at runtime, therefore without the need to write any line of code and with the major advantage that each update to the parameter list is completely transparent to this module. 

\begin{figure*}[!hbt]
\centering
\includegraphics[width=\textwidth]{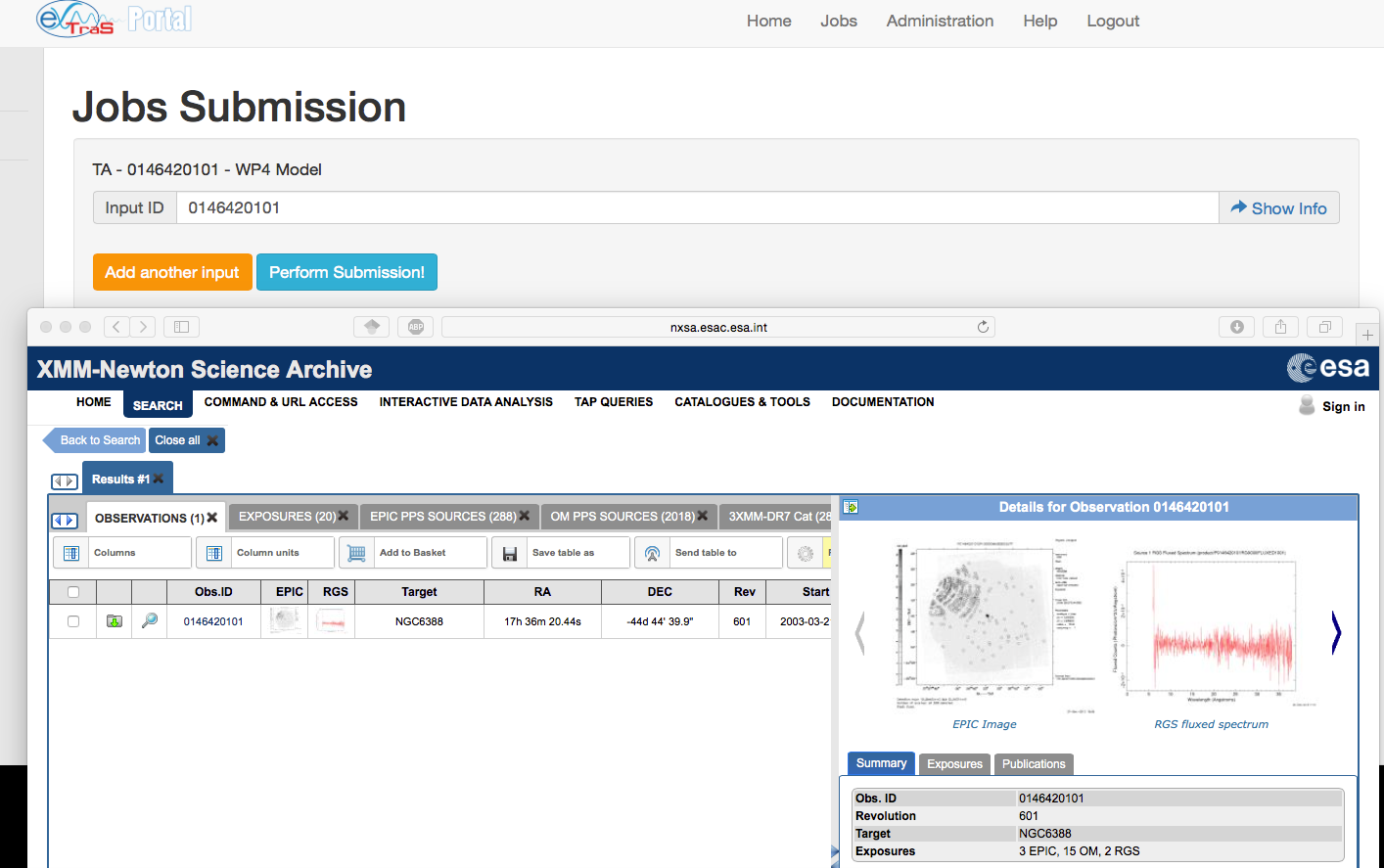}
\caption{
The portal allows to query the XSA for validating the inserted OBSID.
}
\label{fig_sh1}
\end{figure*}

\begin{figure*}[!hbt]
\centering
\includegraphics[width=\textwidth]{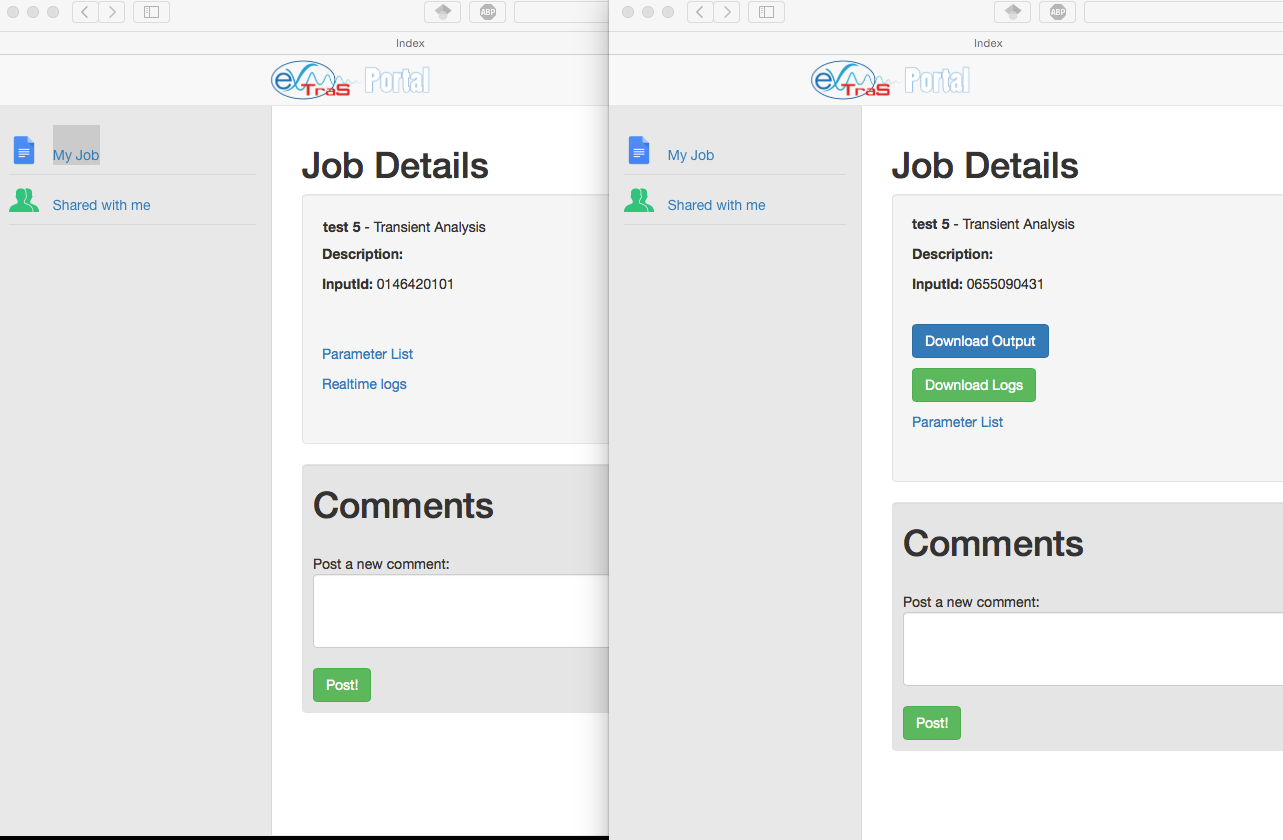}
\caption{The left side of the image shows the interface provided by the FedCloud Submission Handler module during the execution of the job, the right one when it finishes with a success. 
}
\label{fig_sh2}
\end{figure*}

The FedCloud Submission Handler module manages submission of jobs by interacting with a dedicated microservice, described in the following Section,  and provides a full view of the job status, results and logs. In particular the actual submission requires the user specifies as input for the analysis one or more observation identifiers (OBSID) among those contained in the XSA, as shown in Figure \ref{fig_sh1}. Each of them corresponds to a job, therefore an analysis configuration can result in multiple jobs executed at a time.  During the execution the user can monitor the status of the job by means of the real-time log information the software tool provides.
When a job terminates, the FedCloud Submission Handler  provides the possibility to retrieve results and also log informations, as shown in Figure \ref{fig_sh2}.  All the information related to a job (e.g. the configuration parameters, the logs, the results, the ownership/sharing information and possible comments) are stored on the portal database via the Persistence API until it is deleted by the user who owns it. 
It is to note that the FedCloud Submission Handler module can be called also by external systems via an API. In perspective the EXTraS portal will provide the analysis software as services. 

The  EXTraS portal provides two further key features: the ability to share an analysis (i.e. the namelist and possibly the results) and the support for the interaction and discussion (in terms of comments) among the scientists sharing it.
Sharing a job means not only that the experiment results are visible to other users, but also the configuration is shared and can be used as a starting point for re-submitting the experiment on a new set of data. Thus, a job execution can be replicated by other users that can, for example, validate the experiment results or explore the behavior by changing one or a few parameter values. 


\subsection*{Json-GUI}
Json-GUI\footnote{https://github.com/portalTS/json-gui} is a front-end library, developed as a set of reusable AngularJS directive, that allows the dynamic generation of full-featured form-based web interfaces including validation and constraints. 
It can considered a companion tool with respect to PortalTS, but it can be exploited in any AngularJS/Javascript web application.

The defining characteristic of this library is that it simplifies and automatizes the design and the implementation of a standard Web interface, reducing the development time. Starting from a formal JSON configuration describing a list of inputs, this module is able to build a form frame interface, with standard type  but also personalized validation and constraints. 

In details the library provides 7 basic parameter types: integer, float, datetime, text, select, geo-referenced domains and a generic fileupload.
For each parameter, the module allows to define different constraints with different error messages, where each constraint can be constituted by one or more conditions. In such a way it is possible to explicitly support all the HTML 5 input types. 

Moreover, within each condition, the user can compare the value of the parameter with the value of another one, and/or with a static value. Figure \ref{fig_json} shows how it is possible to define an interface starting from a definition of the mandatory selection parameter $TIME\_INTERVAL\_SELECTION$.  We also assume  the previous definition of a parameter named $TIME\_INTERVAL\_SELECTION\_BAYESIAN$. The former value depends on the latter one because the ``dependencies'' property defines a relation between them, and a set of conditions with the  ``isValid'' property that have to be verified. If this condition is not verified, the interface raises an error message. 

Json-GUI is of particular importance when developers have to deploy in the science gateways software modules subject to frequent updates. As stated before, each UI in fact  is dynamically built by the Workflow Configuration module starting from the JSON file corresponding to the kind of analysis a user wants to perform. This means that there is not the need to write any line of code, giving however the possibility to specify personalized behavior and accepted value for each parameter. The ability to modify the interface simply changing a configuration file allows a faster development cycle.

The choice of building such a module from scratch, against using already existing ones, is derived from the need to address some specific requirements. In particular, the way the module is built lets any non-IT scientist to easily design an advanced configuration interface, freeing the IT developer from editing the source code any time a scientist decides to update the parameter list or any other interface element. In fact, Json-GUI comes with some high-level features well suited for the scientific context, that gives scientists the possibility to easily define high level validation and constraints between configuration parameters.

\section{The EGI Computational Infrastructure}
\label{sec4}

The EXTraS Science Gateway relies on the EGI  Cloud Compute service to provide scientists with the possibility to run the EXTraS analysis tools on all the observation data available through the XMM-Newton Science Archive (XSA). 
To allow users from the Astrophysics community to analyse their own data using EXTraS pipelines and guaranteed a reliable services to end-users, the project has signed an agreement (SLA) with three cloud providers of the EGI Federation and created the \emph{extras-fp7.eu} VO. 
In details, the INFN-CATANIA-STACK resource center provides 10 Virtual CPU cores, 40 GB of memory and a scratch storage of 0.6 TB; the RECAS-BARI resource center provides 10 Virtual CPU cores, 40 GB of memory and a scratch storage of 1 TB; the CYFRONET-CLOUD resource center provides 40 Virtual CPU cores, 160 GB of memory and a scratch storage of 100 GB. Presently, we are able to run up to 30 analysis at a time \cite{egisla} by exploiting a IaaS infrastructure. 

The resulting computational infrastructure is depicted in the left side of Figure \ref{fig_arch} and it is composed by four components, the EGI Applications Database (AppDB), the E-Token Server, the CERN Virtual Machine File System infrastructure (CVMFS) and the  cloud providers of the EGI Federation (FedCloud) that actually support the execution of the EXTraS analysis tools. The usage of the components is managed by a microservice developed as part of the EXTraS portal.

\subsection*{AppDB}
The EGI Applications Database is a central marketplace containing computing tools ready to be used on top of the EGI Grid and/or Cloud Infrastructure. In particular the tools can be represented by applications, for which is provided a description and information of the virtual organizations (VO) that can exploit them on a set of computational resources, and virtual appliance, i.e. pre-configured virtual machine images that can be instantiated again by members of one or more VO and using the Cloud infrastructure of  the providers supporting the VOs.

In the first release of the EXTraS portal we adopted the approach of preparing a dedicated virtual appliance for each analysis tool. However we had to modify it on the basis of the characteristics of the EXTraS software tools. 
They rely in fact on the Science Analysis System\footnote{https://www.cosmos.esa.int/web/xmm-newton/what-is-sas} (SAS), a collection of software to analyze the data collected by the XMM-Newton observatory, and HEAsoft\footnote{https://heasarc.nasa.gov/lheasoft/}, a collection of software to manipulate FITS Files and to analyze data for high-energy astronomy. Both these packages are rather complex to install and setup and, more importantly, require about 9 GB of disk space, including calibration files required by SAS.  Considering that every change in the EXTraS software requires to re-upload the entire virtual appliance on the AppDB and that most of the Cloud providers do not allow virtual appliance with a disk larger than 10 GB, we decided to move all the EXTraS-specific software and packages on CVMFS. 

Therefore we use a general-purpose virtual appliance running Ubuntu, which  is started and configured at runtime by using specific contextualization scripts by the microservice described below. In particular the virtual machine instance installs at runtime some general-purpose packages for SAS and Heasoft, some Python-based packages for astrophysics, and mount the CVMFS directory containing the EXTraS software. Moreover, considering also I/O and temporary data for the analysis can be of several GB, the virtual machine uses also a block storage device, that is dynamically created and mounted at runtime by the microservice using the Open Cloud Computing Interface (OCCI).

\subsection*{E-Token Server}
During the aforementioned DRIHM project we realized that it is unfeasible to require all the users of a science gateway to get a digital certificate issued by a recognized Certification Authority (CA).
EGI in fact  relies on a single sign-on mechanism to access the federated services based on X.509 certificates and VO membership. The solution we adopted there and  in the EXTraS portal relies on the use of robot certificates and e-Token servers \cite{etoken}.

Robot certificates have been introduced by the Inter-operable Global Trust Federation (IGTF) in 2010. The formal definition is the following: ``Robots, also known as automated clients, are entities that perform automated tasks without human intervention. Production ICT environments typically support repetitive, ongoing processes - either internal system processes or processes relating to the applications being run (e.g. by a site or by a portal system). These procedures and repetitive processes are typically automated, and generally run using an identity with the necessary privileges to perform their tasks'' \cite{robot}.

In practice, they were introduced to allow users, who cannot get or are not familiar with personal digital certificates, to exploit any distributed infrastructure relying on them in their research activities. The robot certificate is usually associated with a specific application (or function) that the application developer/provider wants to share with all the VO \cite{robot2}. This is exactly the scenario arisen in the EXTraS activities, because portal users are provided with the  possibility to run only pre-defined software tools.

Robot certificates are used, together with one or more MyProxy servers, to generate short lived certificates called proxy certificates. These proxy certificates are then actually used to perform the execution of the software on the resources supporting the VO. An evolution of this concept is represented by the per-user sub-proxy (PUSP), that allows identification of the individual users that operate using a common robot certificate. This is achieved by creating a proxy credential from the robot credential with the proxy certificate containing user-identifying information in its additional proxy CN field. 

To be complaint with the EGI policies, the EXTraS portal has been configured to send a PUSP generation request via its network API. In this request, the portal specifies the ID its associate to the user, which will be used by the e-Token server to generate the short-term proxy certificate that can be used to interact with the extras-fp7.eu VO resources.

\subsection*{CVMFS}
The CERN Virtual Machine File System (CernVM-FS or CVMFS) has been designed as a scalable, reliable and low-maintenance software distribution service with the aim to replace local package managers and shared software areas on cluster file systems. Its original purpose was to support the research activities in High Energy Physics at CERN by deploying software through a a POSIX read-only file system in user space \cite{cvmfs}. With this approach files  are hosted on standard web servers and mounted, uses outgoing HTTP connections only, in the universal namespace \emph{/cvmfs}.
As stated in the documentation, CVMFS focuses specifically on the software use case by means of aggressive caching and reduction of latency. Software in fact usually comprises many small files that are frequently opened and read as a whole. Furthermore, the software use case includes frequent look-ups for files in multiple directories when search paths are examined. Data and meta-data are therefore transferred on demand and data integrity is verified  by cryptographic hashes.

\subsection*{The FedCloud Microservice}

A microservice is a self-contained reusable component that fulfills a specific task. They are loosely coupled components that 
provide better scalability with respect to other solutions \cite{msl}.
For example, if a single microservice of a portal is accessed by many more users than the others,  thus it needs to manage many more connections and activities, it is possible to instantiate it multiple times, performing load balancing without wasting resources, thus scaling at very low granularity. On the contrary, with a traditional single Web Service approach, the whole Web Service requires to be instantiated multiple times, increasing the consumption of partially exploited resources.

Furthermore it is possible to update, improve or correct bugs of each microservice independently from the others. This enables a faster development cycle and simpler analysis, debug and deployment phases for each microservice, improving the maintenance of the whole system and enabling a fast prototyping on new functionalities. 

In the current EXTraS portal the scalability does not represent a key issue, while the independent development cycle does. In particular we designed the usage of microservices for specific computational infrastructure. Their duty in fact is to get the kind of analysis and parameters the user requested, the (set of ) OBSID, and to submit the execution on a specific computational infrastructure.

Each microservice therefore expose two different APIs. The first one is used by a corresponding submission handler, to submit and monitor the experiments. The other one supports the communication with the computing resources used to run the experiments, in order to provide a nearly real-time feedback log to the user and manage the job termination.

In the current production release of the EXTraS portal we support only one computational infrastructure, therefore only one microservice has been deployed. It is responsible to run the job by a proper contextualization of one (or more) instance of the virtual appliance on EGI FedCloud and destroy the virtual machine when the analysis has been executed with success or an error occurs.   

In detail, the FedCloud Submission Handler performs an AJAX call to the microservice providing the user and job ID as stored in the portal repository. The microservice 
a) interacts with the Persistence API to retrieve the parameter list, the kind of analysis to perform and, the list of OBSID to process; b) interacts with the E-Tokern server to get a PUSP; c) selects (presently in a round-robin way) one of the Federated Cloud site supporting the extras-fp7.eu VO; d) runs an instance of the virtual appliance providing a proper contextualization scripts via a OCCI client\footnote{https://github.com/EGI-FCTF/rOCCI-cli}; e) receives updates and log informations by a process local to the virtual machine that provides stdio and stderr in a progressive way via POST requests; f) deletes the virtual machine when the status update notifies the end of the analysis or an error. Result retrieval is performed by the contextualization script via scp.



\section{The Analysis of Highly Variable and Transient Sources}
\label{sec5}

The analysis of highly variable and  transient sources aims at identifying burst-like variability during EPIC observations. It is based on standard source detection algorithm, that are applied to time-resolved images derived from the XMM-Newton observations.  Images are analyzed to identify new point sources that might have brightened considering different energy bands.
For each EPIC observation, a set of images can be obtained by dividing the observing time into sub-exposures corresponding to different time intervals. The time intervals for sub-exposures can
either have fixed duration or they can be defined with a preliminary search for an excess of counts within a small region of the detector in limited time periods. This step of the analysis is performed using a Bayesian Blocks algorithm \cite{scargle} on the events detected in partially overlapping regions having a size comparable to the characteristic dimension of the telescope point spread function.
The main parameters for the transient analysis on one or more observations, named hereafter the experiment, are the choice of instruments (i.e. among the three EPIC cameras), the energy bands, the list of observation identifier(s) and the possible time intervals for source detection.

A full description of the software pipeline is presented in \cite{hb}. 
Different versions of the pipeline were run several times on the full set of EPIC data analyzed to produce the 3XMM-DR5 catalogue and the corresponding results were screened to extend this catalogue with the addition of high confidence transient sources, in part using the science gateway. After a post-processing analysis a final list of 136 new transients was obtained. Most of them (122) were discovered thanks to the Bayesian Blocks algorithm. 
By cross-matching the transient positions with stellar catalogues and inspecting the corresponding optical images, it becomes clear that most transients are very likely stellar flares (including flares from young stellar objects, as shown in \cite{tiengo}).

\section{Conclusions and Future Works}
\label{sec6}
This paper presented an updated release of the EXTraS portal, a science gateway for the astrophysics community devoted to the search and characterization of variable sources in the soft X-ray energy range  by exploiting the XMM-Newton observations. The portal relies on recent general-purpose technologies, architectural patterns and best practices adopted in the development of enterprise web application. 

The EXTraS portal has been validated in December 2016 and officially presented to the scientific community in June 2017 \cite{8b}. Presently, the registered users are a few dozens, but they submitted analysis task for the equivalent of 15,795 processor hours. 

Thanks to this flexible architecture, we can add and manage microservices
in order to improve the resource usage and availability.
To this extent we are working on an extension of PortalTS, named EasyGateway, to be able to interplay with other toolkits. In particular we experimented that the joint use of EasyGateway and Apache Airavata leads to a rich user interface from the Airavata side, with  support for submission on a large set of middleware and queue managers for science gateways relying on PortalTS \cite{dago2}.


The integration of the other analysis tools, both developed in the project and in the scientific community, represents the future direction. We are in particular considering the possibility to provide a remote visualization service based on \cite{vis}, for a better exploitation of experiment results. This feature will be of particular importance for  outreach and students involvement activities of the project and, in perspective, to provide the portal to the large community of citizen scientists interested in the astrophysics research activities.


\section*{Acknowledgment}
EXTraS has received funding from the European Union's 7th Framework Programme  for research, technological development and demonstration under grant agreement no. 607452. This work used the EGI infrastructure with the support of CYFRONET- CLOUD, INFN-CATANIA-STACK and RECAS-BARI. Authors are grateful to Marica Antonacci from RECAS-BARI and Catalin Condurache from Science and Technology Facilities Council for their valuable assistance with the setup of the computing infrastructure.


\begin{thebibliography}{40}



\bibitem{strueder} L. Strueder, U. Briel, K. Dennerl, R. Hartmann, E. Kendziorra, N. Meidinger, E. Pfeffermann, C. Reppin, B. Aschenbach, W. Bornemann, et al., The European Photon Imaging Camera on XMM-Newton: The pn-CCD camera, Astronomy \& Astrophysics, vol. 365, pp. 18, 2001

\bibitem{turner} M. J. L. Turner, A. Abbey, M. Arnaud, M. Balasini, M. Barbera, E. Belsole, P. J. Bennie, J. P. Bernard, G. F. Bignami, M. Boer, et al., The European Photon Imaging Camera on XMM-Newton: The MOS cameras, Astronomy \& Astrophysics, vol. 365, pp. 27, 2001

\bibitem{rosen}
S. R. Rosen, N. A. Webb, M. G. Watson, J. Ballet, D. Barret, V. Braito, F. J. Carrera, M. T. Ceballos, M. Coriat, R. Della Ceca et al.,
The XMM-Newton serendipitous survey. VII. The third XMM-Newton serendipitous source catalogue,
Astronomy \& Astrophysics, vol. 590, pp. A1, 2016.

\bibitem{read}
R. S. Warwick, R. D. Saxton,  A. M. Read, The XMM-Newton slew survey in the 2–10 keV band. Astronomy \& Astrophysics, vol. 548, pp. A99, 2012.

\bibitem{extras1}
A. De Luca, R. Salvaterra, A. Tiengo, D. D'Agostino, M.G. Watson, F. Haberl, J.Wilms, 
Science with the EXTraS Project: Exploring the X-Ray Transient and Variable Sky. The Universe of Digital Sky Surveys, Astrophysics and Space Science Proceedings, vol. 42, pp. 291-295, Springer, 2016.

\bibitem{science}
Israel, G.L., Belfiore, A., Stella, L., Esposito, P., Casella, P., De Luca, A., Marelli, M., Papitto, A., Perri, M., Puccetti, S., Rodríguez Castillo, G.A., Salvetti, D., Tiengo, A., Zampieri, L., D'Agostino, D., Greiner, J., Haberl, F., Novara, G., Salvaterra, R., Turolla, R., Watson, M., Wilms, J., Wolter, A.
An accreting pulsar with extreme properties drives an ultraluminous x-ray source in NGC 5907, Science, 355 (6327), pp. 817-819, 2017.

\bibitem{mus}
R. Mushotzky, What do Astronomers Really Want? Astronomical Data Analysis Software and Systems (ADASS XX) , ASP Conf. Proc., vol. 442, p.235-241, 2011.

\bibitem{drihm1}
D. D'Agostino, E. Danovaro, A. Clematis, L. Roverelli, G. Zereik, A. Parodi, A. Galizia,
Lessons learned implementing a science gateway for hydro-meteorological research. 
Concurrency and Computation: Practice and Experience, vol. 8, n. 7, pp. 2014-2023, 2016.

\bibitem{drihm2}
A. Parodi, D. Kranzlmueller, A. Clematis, E. Danovaro, A. Galizia, L. Garrote, M. C. Llasat, O. Caumont, E. Richard, Q. Harpham, F. Siccardi, L. Ferraris, N. Rebora, F. Delogu, E. Fiori, L. Molini, E. Foufoula-Georgiou, D. D'Agostino,
DRIHM(2US): an e-Science environment for hydro-meteorological research on high impact weather events. Bulletin of the American Meteorological Society, online, 2017.


\bibitem{iwsg}
D. D'Agostino, L. Roverelli, G. Zereik, A. De Luca, R. Salvaterra, A. Belfiore, G. Lisini, G. Novara, A. Tiengo, A Microservice-Based Portal for X-ray Transient and Variable Sources. Proceedings of the 8th international workshop on Science gateways (IWSG 2016), CEUR-WS.org/Vol-1871, 2017.

\bibitem{fedcloud}
D. C. H. Wallom, M. Turilli, M. Drescher, D. Scardaci and S. Newhouse, Federating Infrastructure as a Service Cloud Computing Systems to Create a Uniform E-infrastructure for Research, Proceedings of the IEEE 11th International Conference on e-Science, pp. 155-164, 2015.

\bibitem{AT}
J. Ruiz, J. Garrido, J. Santander-Vela, S. Sanchez-Exposito, L. Verdes-
Montenegro Astrotaverna-building workflows with virtual observatory
services Astronomy and Computing, 78, pp. 3-11, 2014.

\bibitem{AT2} 
Castelli, G., Taffoni, G., Sciacca, E., Becciani, U., Costa, A., Krokos, M., Pasian F, Vuerli, C., VO-compliant workflows and science gateways,
Astronomy and Computing, 11, pp. 102-108, 2015.

\bibitem{sg2}
Kacsuk P (ed), Science Gateways for Distributed Computing Infrastruc-
tures, ISBN 978-3-319-11268-8, 2014.

\bibitem{sg}
K.A. Lawrence, M. Zentner, N. Wilkins-Diehr, J.A. Wernert, M. Pierce, S. Marru, S. Michael, Science gateways today and tomorrow: Positive perspectives of nearly 5,000 members of the research community. Concurrency and Computation: Practice and Experience , Vol. 27, Iss. 16, pp. 4252-4268, 2015.

\bibitem{visivo}
U. Becciani, E. Sciacca, A. Costa, P. Massimino, C. Pistagna, S. Riggi, F. Vitello, C. Petta, M . Bandieramonte, M. Krokos, Science gateway technologies for the astrophysics community, Concurrency and Computation: Practice and Experience, 27, pp. 306-327, 2015.

\bibitem{becciani2}
A. Costa, P. Massimino, M. Bandieramonte, U. Becciani, M. Krokos, C. Pistagna, S. Riggi, E. Sciacca, F. Vitello, An Innovative Science Gateway for the Cherenkov Telescope Array. J Grid Computing vol. 13, issue 4, pp. 547-559, 2015. 

\bibitem{astro3}
S. Sanchez-Exposito, P. Martin, J. E. Ruiz, L. Verdes-Montenegro, J. Garrido, R. Sirvent, A. Ruiz Falco, R. M. Badia, Web Services as Building Blocks for Science Gateways in Astrophysics.  J Grid Computing vol. 14, issue 4, pp. 673-685, 2016. 


\bibitem{guse}
\'A. Balasko, Z. Farkas, P. Kacsuk, Building Science Gateway by Utilizing the Generic WS-PGRADE/gUSE Workflow System. 
Computer Science, vol. 14, n. 2, pp. 307-325, 2013.

 \bibitem{airavata}
M. E. Pierce, S. Marru, L. Gunathilake, D. Kushan Wijeratne, R. Singh, C. Wimalasena, S. Ratnayaka, S. Pamidighantam, Apache Airavata: design and directions of a science gateway framework. Concurrency Computat.: Pract. Exper, 27, pp. 4282-4291, 2015.

\bibitem{agave}
R. Dooley, M. Vaughn, D. Stanzione, S. Terry, E. Skidmore, Software-as-a-Service: The iPlant Foundation API, 5th IEEE Workshop on Many-Task
Computing on Grids and Supercomputers (MTAGS), IEEE, 2012.

\bibitem{hubzero}
M. McLennan, R. Kennell, HUBzero: A Platform for Dissemination and Collaboration in Computational Science and Engineering, Computing in Science and Engineering, 12(2), pp. 48-52, 2010.

\bibitem{jp}
Kluyver, T., Ragan-Kelley, B., Pérez, F., Granger, B. E., Bussonnier, M., Frederic, J., ... \& Ivanov, P. (2016, May). Jupyter Notebooks-a publishing format for reproducible computational workflows. In ELPUB (pp. 87-90). Proceedings of the 20th International Conference on Electronic Publishing

\bibitem{jph}
Zonca, A. (2016). Jupyter notebooks in science gateways (No. e2577v2). PeerJ Preprints.


\bibitem{dago}
D. D'Agostino, E. Danovaro, A. Clematis, L. Roverelli, G. Zereik, A. Parodi, A. Galizia, 
From lesson learned to the refactoring of the DRIHM science gateway for hydro-meteorological research,
Journal of Grid and Utility Computing, vol. 14, issue 4, pp 575–588, 2016



\bibitem{egisla}
EGI-EXTraS Service Level Agreement. http://goo.gl/TqUm9Y

\bibitem{etoken}
Valeria Ardizzone, Roberto Barbera, Antonio Calanducci, Marco Fargetta, E. Ingrà, Ivan Porro, Giuseppe La Rocca, Salvatore Monforte, R. Ricceri, Riccardo Rotondo, Diego Scardaci, Andrea Schenone: The DECIDE Science Gateway. Journal of Grid Computing 10(4): 689-707 (2012)


\bibitem{robot}
EUGridPMA, Guideline on IGTF Approved Robots, OID 1.2.840.113612.5.4.1.1.1.6, 2014. http://goo.gl/p42E1X 


\bibitem{robot2}
R. Barbera, G. Andronico, G. Donvito, A. Falzone, J. J. Keijser, G. La Rocca, L. Milanesi, G. P. Maggi andS. Vicario
A grid portal with robot certificates for bioinformatics phylogenetic analyses
Concurrency and Computation: Practice and Experience
Volume 23, Issue 3, pages 246-255,  2011



\bibitem{cvmfs}
 J. Blomer, P. Buncic, R. Meusel, G. Ganis, I. Sfiligoi, D. Thain, The Evolution of Global Scale Filesystems for Scientific Software Distribution. Computing in Science \& Engineering, Vol. 17, Iss. 6, pp. 61-71, 2015

\bibitem{msl}
S. Newman, Sam, Building Microservices,  O'Reilly Media, 2015.

\bibitem{scargle} J. D. Scargle, J. P. Norris, B. Jackson, J. Chiang, J., Studies in Astronomical Time Series Analysis. VI. Bayesian Block Representations, The Astrophysical Journal, vol. 764, pp. 167, 2013

\bibitem{hb}
Handbook of transient analysis software, deliverable 4.8 of the EXTras project.
http://www.extras-fp7.eu/images/final/WP4\_pipeline\_transient.pdf

\bibitem{tiengo}
D. Pizzocaro, B. Stelzer, R. Paladini, A. Tiengo, G. Lisini, G. Novara, G. Vianello, A. Belfiore, M. Marelli, D. Salvetti, I. Pillitteri, S. Sciortino, D. D'Agostino, F. Haberl, M. Watson, J. Wilms, R. Salvaterra, A. De Luca, 
Results from DROXO IV. EXTraS discovery of an X-ray flare from the Class I protostar candidate ISO-Oph 85.
Astronomy and Astrophysics, Vol. 587, pp. A36, 2016.

\bibitem{8b}
A. De Luca, R. Salvaterra, A. Tiengo, D. D'Agostino, M.G. Watson, F. Haberl, J. Wilms (2017) EXTraS: Exploring the X-Ray Transient and Variable Sky. The X-ray Universe 2017 Symposium abstract book, 197.

\bibitem{dago2}
A. Galizia, L. Roverelli, G. Zereik, E. Danovaro, A. Clematis, D. D'Agostino (2017).  Using Apache Airavata and EasyGateway for the creation of complex science gateway front-end. Future Generation Computing Systems, online, DOI: 10.1016/j.future.2017.11.033 .

\bibitem{vis}
Handbook of the screening software, deliverable 6.9 of the EXTras project. http://www.extras-fp7.eu/images/final/WP6\_screening.pdf


\end{thebibliography}
\end{document}